\documentclass[journal]{IEEEtran}

\usepackage{booktabs}
\usepackage{bm}
\usepackage{graphicx}
\usepackage[cmex10]{amsmath}
\usepackage{cite}
\usepackage{amssymb}
\usepackage{subfigure}
\usepackage{extarrows}
\usepackage[letterspace = -2]{microtype}
\usepackage{algorithm}
\usepackage{algpseudocode}
\usepackage{float}
\usepackage{upgreek}
\usepackage{color}

\usepackage[absolute,overlay]{textpos}

\newfloat{algorithm}{t}{lop}

\TPMargin{26pt}

\newcommand{\copyrightstatement}{
    \begin{textblock}{0.9}(0.05,0.92)    
         \noindent
         \footnotesize
         \textbf{$\copyright$ 2021 IEEE. Personal use of this material is permitted. Permission from IEEE must be obtained for all other uses, in any current or future media, including reprinting/republishing this material for advertising or promotional purposes, creating new collective works, for resale or redistribution to servers or lists, or reuse of any copyrighted component of this work in other works. DOI: 10.1109/LCOMM.2021.3074756}
    \end{textblock}
}

\begin{document}
\title{\huge Dynamic Resource Configuration for Low-Power IoT Networks: A Multi-Objective Reinforcement Learning Method
}

\author{
Yang~Huang,~\IEEEmembership{Member,~IEEE},
Caiyong Hao,
Yijie Mao,~\IEEEmembership{Member,~IEEE}
and Fuhui Zhou,~\IEEEmembership{Senior Member,~IEEE}

\thanks{This work was partially supported by the National Natural Science Foundation (NSF) of China under Grant 61901216, U2001210, 62071223 and 62031012, the NSF of Jiangsu Province under Grant BK20190400, the open research fund of National Mobile Communications Research Laboratory, Southeast University (No. 2020D08). (Corresponding author: Caiyong Hao)}
\thanks{Y. Huang and F. Zhou are with the Key Laboratory of Dynamic Cognitive System of Electromagnetic Spectrum Space, Ministry of Industry and Information Technology, Nanjing University of Aeronautics and Astronautics, Nanjing, 210016, China (email: \{yang.huang.ceie, zhoufuhui\}@nuaa.edu.cn). Y. Huang is also with the National Mobile Communications Research Laboratory, Southeast University, Nanjing, 210016, China.

C. Hao is with the Shenzhen Station of State Radio Monitoring Center, Shenzhen, 518000, China (e-mail: hao.c.y@srrc.org.cn). He is also with the School of Electronic Information, Wuhan University, Wuhan, 430072, China.

Y. Mao is with the Department of Electrical and Electronic Engineering, Imperial College London, London SW7 2AZ, United Kingdom (e-mail: y.mao16@imperial.ac.uk).}

}

\maketitle

\vspace{-0.2cm}
\begin{abstract}
Considering grant-free transmissions in low-power IoT networks with unknown time-frequency distribution of interference, we address the problem of Dynamic Resource Configuration (DRC), which amounts to a Markov decision process. Unfortunately, off-the-shelf methods based on single-objective reinforcement learning cannot guarantee energy-efficient transmission, especially when all frequency-domain channels in a time interval are interfered. Therefore, we propose a novel DRC scheme where configuration policies are optimized with a Multi-Objective Reinforcement Learning (MORL) framework. Numerical results show that the average decision error rate achieved by the MORL-based DRC can be even less than 12\% of that yielded by the conventional R-learning-based approach.
\end{abstract}

\begin{IEEEkeywords}
IoT networks, multi-objective reinforcement learning, grant-free, spectrum sharing.
\end{IEEEkeywords}

\copyrightstatement

\section{Introduction}
Internet-of-Things (IoT) devices are envisioned to account for 50 percent (14.7 billion) of the global connected devices by 2023, among which nearly one third is expected to be wireless IoT devices \cite{Cisco2023}. In order to accommodate spectrum utilization for such ubiquitous but massive IoT devices, spectrum sharing is expected to be a promising solution\cite{ZLX19}.
However, since IoT communications are dominated by uplink (UL) transmissions of short packets, directly applying a grant-based radio access, which is common in cellular networks, can introduce excessive signaling overhead for short packets, especially in the presence of massive connections\cite{MAYC18}.
Fortunately, it was reported that Grant-Free (GF) UL transmissions \cite{JABPMKM17} can provide IoT devices with energy-efficient communications \cite{MAYC18}.
Although GF scheduling can assign a certain UE dedicated or shared resources\cite{3GPP_R11705654}, recent studies\cite{MAYC18, KSP20} focused on contention-based transmission schemes with shared resources.

\begin{figure}[t]
\centering
\includegraphics[width = 2.8in]{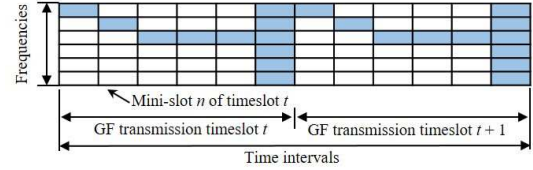}
\caption{The time-frequency resources for GF uplink transmissions. The rectangles in blue mean that the corresponding frequencies and mini-slots are occupied by interference from other wireless devices. Frequency-domain resources in mini-slot 6 of every timeslot are totally occupied by interference.}
\label{Fig_GFtimeDomainStruct}
\end{figure}
On the contrary, following legacy protocols especially specifications related to industrial IoT\cite{3GPP_TR21916}, this paper considers GF scheduling with dedicated resources and develops a Dynamic Resource Configuration (DRC) scheme, rather than always pre-allocating \emph{periodic} radio resources as in semi-persistent scheduling\cite{JABPMKM17}, to improve spectrum usage.
Specifically, in the studied scenario, an Access Point (AP) configures low-power IoT devices (i.e. UEs) with time-frequency resources (as shown in Fig. \ref{Fig_GFtimeDomainStruct}) by RRC (or with L1) signaling \cite{3GPP_TR21916} prior to each GF UL transmission timeslot, without the knowledge of interference pattern (i.e. the time-frequency distribution of interference from other wireless devices). Moreover, due to the limited energy supply and signal-processing capability at the UEs\cite{ZLX19}, the AP cannot acquire exact channel state information.
Therefore, intuitively, in order to optimize the resource configuration policy and therefore guarantee energy-efficient transmission, the network or AP has to perform sequential decision-making with uncertain interference pattern and channel responses, according to immediate rewards of the amount of successfully received data (i.e. normalized throughput) and the observed spectrum utilization state.
Such a problem boils down to a Markov Decision Process (MDP) \cite{Powell07}, which is reminiscent of anti-jamming communications with Single-Objective (SO) Reinforcement Learning (RL)\cite{LXJWA18}. However, this approach is unable to guarantee energy efficiency in the scenario where frequency resources in certain mini-slots are totally occupied by interference. Intrinsically, regardless of transmitting in such mini-slots or not, the action values obtained by SORL for different actions remain unchanged.

Therefore, in order to maximize the long-term average normalized throughput and simultaneously minimize the long-term average energy consumption (i.e. the number of utilized resource blocks), an approach based on Multi-Objective Reinforcement Learning (MORL) is necessary. Recently, RL has been exploited to optimize radio parameters\cite{Ferreira16} or beam selection\cite{HZLLWG20}, so as to maximize/minimize multiple goals. However, the method in \cite{Ferreira16} can cause ambiguity in selecting the best action due to the utilization of a synthetic value function, while the MORL framework \cite{LXH15} utilized in \cite{HZLLWG20} is inapplicable to an MDP with average cost\cite{SB98}. Therefore, we propose a novel MORL-based DRC scheme, by integrating the weighted-sum framework for MORL \cite{LXH15,NT05} with R-learning\cite{SB98}. Meanwhile, a series of techniques are developed to tackle the high-dimensional state/action space, reduce the space complexity of algorithm and avoid the strictly suboptimal solution resulting from the standard $\epsilon$-greedy strategy.
Numerical results reveal that thanks to the multi-objective framework, the proposed scheme can significantly outperform the conventional RL, achieving much lower decision error rates, especially in the scenario where all the frequency-domain channels are occupied. \emph{Notations}: Matrices and vectors are in bold capital and bold lower cases, respectively; {\small$(\cdot)^T$}, {\small$(\cdot)^\star$}, {\small$\|\cdot\|$}, {\small$|\cdot|$}, {\small$\mathcal{E}\{\cdot\}$} and {\small$\text{Card}\left(\mathcal{A}\right)$} represent the transpose, optimal solution, 2-norm, absolute value, expectation and the cardinality of the set {\small$\mathcal{A}$}, respectively; {\small$\lfloor a \rfloor$} (or {\small$\lceil a \rceil$}) rounds $a$ to the nearest integer less (or greater) than or equal to $a$; {\small$\text{cat}(\mathbf{A},\mathbf{B})$} concatenates {\small$\mathbf{B}$} vertically to the end of {\small$\mathbf{A}$}.

\section{Preliminaries}
\label{SecPreliminaries}
\subsubsection{Signal Model}
\label{SecSignalModel}
The studied UL transmission system consists of an AP and $K_U$ UEs. In each time interval $l$, a certain UE $i \in \{1, \dots, K_U\}$ can deliver an information signal $x_{U,i,l}$ (for $\mathcal{E}\{|x_{U,i,l}|^2\} = p_{U,i}$) to the AP at a frequency of $f_m$ for $m \in \{1,\ldots M\}$. Therefore, let $\alpha_{i, m, l} = 1$ (or $0$) indicate frequency $m$ being (or not) utilized by UE $i$ in time interval $l$, given $i$ and $l$, $\alpha_{i, m, l}$ satisfies
\begin{equation}
\label{EqFreqOccupInd_SU}
\textstyle{\sum_{m=1}^{M}}\alpha_{i, m, l} \leq 1\,.
\end{equation}
Moreover, frequency $f_m$ can be exclusively occupied by the transmission of UE $i$ and $K_U < M$. Namely, signals transmitted by the $K_U$ UEs do not interfere with each other. Hence, given $m$ and $l$, $\alpha_{i, m, l}$ satisfies
\begin{equation}
\label{EqFreqOccupInd_MU}
\textstyle{\sum_{i=1}^{K_U}}\alpha_{i, m, l} \leq 1\,.
\end{equation}
In the meanwhile, by designating the interference signal emitted by interferer $j$ as $x_{I,j,l}$ for $j \in \{1,\ldots K_I\}$ and $\mathcal{E}\{|x_{I,j,l}|^2\} = p_{I,j}$, the baseband signal $y_{m,l}$ received by the AP at frequency $m$ in time interval $l$ can be expressed as
$y_{m,l}\! =\! \sum_{i = 1}^{K_U} \alpha_{i, m, l} h_{i, m, l} x_{U,i,l} \! + \! \sum_{j = 1}^{K_I} \beta_{j, m, l} g_{j, m, l} x_{I,j,l} \! + \! n_{\text{AP}}$,
where $h_{i, m, l}$ and $g_{j, m, l}$ respectively represent the channel gain between the AP and UE $i$ and that between the AP and interferer $j$ at frequency $m$ in time interval $l$; the random variable $n_\text{AP}\sim \mathcal{CN}(0, \sigma_n^2)$ stands for the noise at frequency $m$ in time interval $l$. Similarly to $\alpha_{i, m, l}$, $\beta_{j, m, l} \in \{0,1\}$ in the above equation indicates whether frequency $m$ is exploited by interferer $j$ for transmission in time interval $l$.

\subsubsection{Grant-Free UL Transmissions}
\label{SecGFtransmission}
As shown in Fig. \ref{Fig_GFtimeDomainStruct}, a timeslot spans several time intervals, such that a UE can be configured to transmit in multiple time intervals. Then, in the subsequent timeslot, the UEs deliver information through the configured time-frequency channels. For generality, we consider that each timeslot consists of $N$ mini-slots, where the duration of each mini-slot is identical to that of a time interval. Let $n \in \{1, \ldots, N\}$ and $t$ respectively denote the index of a mini-slot in a timeslot and the index of a timeslot, $n$ and $t$ satisfy $n = l - N\cdot \lfloor l/N \rfloor$ and $t = \lceil l/N \rceil$, which indicates the relation between time intervals and the timeslot for GF transmission.
Therefore, with GF transmission, the signal at frequency $m$ received by the AP in mini-slot $n$ of timeslot $t$ is formulated by recasting $y_{m,l}$ as $y_{m,n,t} = \sum_{i = 1}^{K_U} \alpha_{i, m, n, t} h_{i, m, n, t} x_{U, i, n, t} + \sum_{j = 1}^{K_I} \beta_{j, m, n, t} g_{j, m, n, t} x_{I, j, n, t} + n_\text{AP}$, where $\alpha_{i, m, n, t}$, satisfying (\ref{EqFreqOccupInd_SU}) and (\ref{EqFreqOccupInd_MU}), indicates the time-frequency resource configuration for UE $i$ and is determined by the AP (or the network). In contrast, $\beta_{j, m, n, t}$ depends on the time-frequency interference distribution, which is unknown to the AP. When frequency resources in a mini-slot (for given $n$ and $t$) are totally occupied by interferers, $\beta_{j, m, n, t} = 1 \quad \forall m$.

The average power of the signal at frequency $m$ measured at the AP in mini-slot $n$ of timeslot $t$ can be obtained as $\rho_{m,n,t} = \mathcal{E}\{y_{m,n,t}\} = \textstyle{\sum_{i = 1}^{K_U}} \alpha_{i, m, n, t} |h_{i, m, n, t}|^2 p_{U,i} + \textstyle{\sum_{j = 1}^{K_I}} \beta_{j, m, n, t} |g_{j, m, n, t}|^2 p_{I,j} + \sigma_n^2.$
We assume that the AP cannot acquire exact channel state information on $h_{i, m, n, t}$ \cite{ZLX19} and has no knowledge of the dynamic interference.
The receive Signal-to-Interference-plus-Noise Ratio (SINR) with respect to (w.r.t.) the transmission of UE $i$ in mini-slot $n$ of timeslot $t$ can be obtained as
\begin{equation}
\label{EqSINR}
\gamma_{i,n,t} = \frac{ \textstyle{\sum_{m=1}^M} \alpha_{i, m, n, t} |h_{i, m, n, t}|^2 p_{U,i}}{ \textstyle{\sum_{m=1}^M} \textstyle{\sum_{j = 1}^{K_I}} \beta_{j, m, n, t} |g_{j, m, n, t}|^2 p_{I,j} + \sigma_n^2 }\,.
\end{equation}
Given an SINR threshold $\gamma_0$, if $\gamma_{i,n,t} \geq \gamma_0$, the transmitted signal can be decoded. Thus, the normalized throughput for UE $i$ in mini-slot $n$ of timeslot $t$ can be defined as
$u_{i,n,t} \triangleq \delta(\gamma_{i,n,t} - \gamma_0)$,
where $\delta(\gamma_{i,n,t} - \gamma_0)= 1$, if $\gamma_{i,n,t} \geq \gamma_0$; otherwise, $\delta(\gamma_{i,n,t} - \gamma_0)= 0$.

\section{Problem Formulation}
\label{SecProblemForm}
This paper addresses the problem of dynamic resource configuration. It can be inferred from (\ref{EqSINR}) that the key is to dynamically adjust the resource configuration $\alpha_{i, m, n, t}$ $\forall n,t$ but avoid the co-channel interference.
However, according to Section \ref{SecGFtransmission}, in a certain time interval $l$ (which corresponds to a mini-slot in the GF transmission), the AP has no knowledge of the average power $\sum_{m=1}^M \sum_{j = 1}^{K_I} \beta_{j, m, l+1} |g_{j, m, l+1}|^2 p_{I,j}$ or the frequency occupation $\beta_{j, m, l+1}$ of the interference in the next time interval $l+1$.
Intuitively, in order to determine the resource configuration $\alpha_{i, m, n, t+1}$ $\forall n$ for UEs' transmission in timeslot $t+1$, the AP has to rely on the observed spectrum utilization state i.e. $\rho_{m,n,t}$ $\forall m,n$ and the normalized throughput $u_{i,n,t}$ $\forall i,n$ in timeslot $t$.
Therefore, the design problem boils down to an MDP.

In order to formulate the MDP, we designate the state space $\mathcal{S}$ as a set that collects states of the spectrum environment in a timeslot, while the action space $\mathcal{A}$ is a set that collects all possible time-frequency resource configuration in a timeslot.
Specifically, in timeslot $t$, the observed spectrum utilization state $\mathbf{s}_t \in \mathcal{S}$ can be expressed as $\mathbf{s}_t = [\mathbf{s}^T_{t,1}, \ldots, \mathbf{s}^T_{t,n}, \ldots \mathbf{s}^T_{t,N}]^T$, where $\mathbf{s}_{t,n} = [\rho_{1,n,t},\ldots,\rho_{M,n,t}]^T$ collects the spectrum utilization situations in mini-slot $n$.
The resource configuration action $\mathbf{a}_t \in \mathcal{A}$ can be written as $\mathbf{a}_t = [\mathbf{a}^T_{t,1}, \ldots, \mathbf{a}^T_{t,n}, \ldots \mathbf{a}^T_{t,N}]^T$, where $\mathbf{a}_{t,n}$ collects the frequency resource configurations for the $K_U$ users in mini-slot $n$; we define that $\mathbf{a}_{t,n} \triangleq \text{vec}(\mathbf{A}_{t,n})$, and $\mathbf{A}_{t,n}$ is an $M$-by-$K_U$ matrix, where each entry $[\mathbf{A}_{t,n}]_{m,i} = \alpha_{i, m, n, t}$. The dynamics is nothing else than the transition probability $P(\mathbf{s}_{t+1} = s^\prime | \mathbf{s}_t = s, \mathbf{a}_t= a)$ for $s \in \mathcal{S}$ and $a \in \mathcal{A}$, which is however unknown to the AP, as well as the UEs. Due to this, intuitively, the formulated MDP needs to be solved by RL.

Frequency-domain resources in certain mini-slots can be totally occupied by interference signals, In order to reduce energy consumption, the UEs should not be configured to transmit during such mini-slots.
Unfortunately, such an issue cannot be handled by SORL which involves a scalar immediate reward\cite{LXJWA18}.
Therefore, we formulate the immediate reward as a vector which is given by $\mathbf{r}_t = [R_t, - P_t]^T$, where
$R_t = \sum_{n=1}^N \sum_{i=1}^{K_U} u_{i,n,t}$ and $P_t = \sum_{n=1}^N \sum_{i=1}^{K_U} \sum_{m=1}^M \alpha_{i, m, n, t}$
respectively evaluate the overall normalized throughput and the energy consumption in a timeslot.
In order to improve the expected long-term average throughput as well as the energy efficiency, the average reward w.r.t. $R_t$ and $-P_t$ can be respectively obtained as
{\small$\bar{R} = \lim_{T \rightarrow \infty} \sup \frac{1}{T}\mathcal{E}\left\{\sum_{t=0}^{T-1}R_t\right\}$} and {\small$\bar{P} = \lim_{T \rightarrow \infty} \sup \frac{1}{T}\mathcal{E}\left\{-\sum_{t=0}^{T-1}P_t\right\}$},
which can be further collected in $\bar{\mathbf{r}} \triangleq [\bar{R}, \bar{P}]^T$.
Hence, by defining a weight vector $\mathbf{w} = [w_R, w_P]^T$ which indicates the network or the AP's preferences between different objectives, the optimization of the configuration policy $\pi$, which is a deterministic policy given by $\pi: \mathcal{S} \rightarrow \mathcal{A}$, can be formulated as
\begin{equation}
\label{EqProbOptimPi}
\pi = \arg \max_\pi \left\{ \mathbf{w}^T \bar{\mathbf{r} }\right\}\,.
\end{equation}

\section{MORL-Based Dynamic Resource Configuration}
\label{SecDynamicRscConfigMORL}
As depicted in Fig. \ref{Fig_GFtimeDomainStruct}, both the observation of the spectrum utilization states and the decision of actions span several mini-slots. This contributes to high-dimensional state space $\mathcal{S}$ and action space $\mathcal{A}$, which make solving the MDP problem (\ref{EqProbOptimPi}) suffer from curses of dimensionality \cite{Powell07}.
Fortunately, due to the fact that the events of spectrum utilization and resource configuration in different mini-slots are statistically independent, the MDP formulated in Section \ref{SecProblemForm} can be decomposed into $N$ MDPs.
In the MDP for mini-slot $n$, the state space and the action space can be respectively recast as $\mathcal{S}_n$ and $\mathcal{A}_n$ for $\mathcal{S} = \cup_{n=1}^N \mathcal{S}_n$ and $\mathcal{A} = \cup_{n=1}^N \mathcal{A}_n$; the immediate reward vector is $\mathbf{r}_{t,n} = [R_{t,n}, - P_{t,n}]^T$, where $R_{t,n} = \sum_{i=1}^{K_U} u_{i,n,t}$ and $P_{t,n} = \sum_{i=1}^{K_U} \sum_{m=1}^M \alpha_{i, m, n, t}$ represent the reward scalars w.r.t. the normalized throughput and energy consumption, respectively. Hence, by defining {\small$\bar{R}_n = \lim_{T \rightarrow \infty} \sup \frac{1}{T} \mathcal{E} \left\{\sum_{t=0}^{T-1} \! R_{t,n}\right\}$}, {\small$\bar{P}_n = \lim_{T \rightarrow \infty} \sup \frac{1}{T} \mathcal{E}\left\{-\sum_{t=0}^{T-1}P_{t,n}\right\}$} and $\bar{\mathbf{r}}_n = [\bar{R}_n, \bar{P}_n]^T$, the $n$\,th subproblem of problem (\ref{EqProbOptimPi}) can be cast as
\begin{equation}
\label{EqSubProbOptimPi_n}
\pi_n = \arg \max_{\pi_n} \left\{ \mathbf{w}^T \bar{\mathbf{r}}_n \right\}\,,
\end{equation}
where $\pi_n: \mathcal{S}_n \rightarrow \mathcal{A}_n$.

\begin{algorithm}
{\small
\caption{MORL-Based Dynamic Resource Configuration}\label{AlgMORL}
\begin{algorithmic}[1]
\Statex \textbf{Initialization:} Set $t=0$. Initialize $\mathbf{s}_{0,n}$ (and the corresponding quantized version $\mathbf{s}_{0,n}^\prime$).  Set $\mathcal{S}_n = \mathbf{s}_{0,n}^\prime$, $\mathcal{M}_{R,n} = q_{0,R} \cdot \mathbf{1}_{1\times \text{card}(\mathcal{A}_n)}$ and $\mathcal{M}_{P,n} = q_{0,P} \cdot \mathbf{1}_{1\times \text{card}(\mathcal{A}_n)}$,  $\forall n\in\{1,\ldots,N\}$.
\Repeat
    \For{$n = 1,\ldots,N$}\! \Comment{Resource configuration at the AP}
        \State Generate a random number $\epsilon_x \sim \mathcal{U}(0,1)$;
        \If{$\epsilon_x \geq \epsilon$}
            \State Given $\mathbf{s}_{t,n}^\prime$, compute $\mathbf{a}_n^\star$ by solving (\ref{EqOptimAction});
        \Else
            \State Given $\mathbf{s}_{t,n}^\prime$, obtain $\mathbf{a}_n^\star$ by randomly picking an action $\mathbf{a}_n \! \in \! \{\mathbf{a}_n | \mathbf{w}^T \mathbf{q}_n (\mathbf{s}_{n}^\prime, \mathbf{a}_n) \! = \! w_R \! \cdot \! q_{0,R} \! + \! w_P \! \cdot \! q_{0,P} , \forall \mathbf{a}_n \! \in \! \mathcal{A}_n \}$;
        \EndIf
        \State For mini-slot $n$, the configuration is $\mathbf{a}_{t,n} = \mathbf{a}_n^\star$;
    \EndFor
    \State The UEs perform UL transmission with the resource configuration $\mathbf{a}_t$; the AP achieves the immediate reward $\mathbf{r}_{t,n} \, \forall t$ and observes the spectrum utilization state $\mathbf{s}_{t+1}$;
    \For{$n = 1,\ldots,N$} \Comment{Check if $\mathbf{s}_{t+1}$ is a new state}
        \State Quantize $\mathbf{s}_{t+1,n}$, yielding $\mathbf{s}_{t+1,n}^\prime$;
        \If{$\|\mathbf{s}_{t+1,n}^\prime - \mathbf{s}_{n}^\prime\|/\|\mathbf{s}_{n}^\prime\| > \eta \quad \forall \mathbf{s}_{n}^\prime \in \mathcal{S}_n$}
            \State Update $\mathcal{S}_n$ by performing (\ref{EqStateSpaceComb}) for $t^\prime = t+1$;
            \State Update $\mathcal{M}_{R,n}$ and $\mathcal{M}_{P,n}$ by performing (\ref{EqLookupTableCombIniVec});
        \EndIf
    \EndFor
    \For{$n = 1,\ldots,N$}
       \State Update $\mathbf{q}_n^{(t+1)}(\mathbf{s}_{t,n}^\prime, \mathbf{a}_{t,n})$ by performing (\ref{EqActValUpdate});
       \If{$\epsilon_x \geq \epsilon$}
            \State Update $\bar{\mathbf{r}}_n^{(t+1)}$ by performing (\ref{EqAvgRewardUpdate});
        \EndIf
    \EndFor
    \State $t = t+1$;
\Until{\text{Stopping criteria}}
\end{algorithmic}
}
\end{algorithm}

We now solve problem (\ref{EqSubProbOptimPi_n}) and propose a DRC scheme based on MORL \cite{NT05} and the R-learning algorithm for MDP with average cost\cite{Mahadevan96, SB98}.
Once observing the spectrum utilization state in mini-slot $n$ at the AP, each entry in the spectrum utilization state vector $\mathbf{s}_{t,n}$ is quantized, and the quantized spectrum utilization state vector is designated as $\mathbf{s}_{t,n}^\prime$.
We define that in mini-slot $n$ of timeslot $t$, the action values (w.r.t. the normalized throughput and the energy consumption) achieved by taking action $\mathbf{a}_{t,n}$ in state $\mathbf{s}_{t,n}^\prime$  can be obtained from action value functions $\mathcal{Q}_{R,n}(\mathbf{s}_{t,n}^\prime, \mathbf{a}_{t,n})$ and $\mathcal{Q}_{P,n}(\mathbf{s}_{t,n}^\prime, \mathbf{a}_{t,n})$, respectively.
Basically, in this paper, given $\mathbf{s}_{t,n}^\prime$, we obtain action values from $\mathcal{Q}_{R,n}(\mathbf{s}_{t,n}^\prime, \mathbf{a}_{t,n})$ and $\mathcal{Q}_{P,n}(\mathbf{s}_{t,n}^\prime, \mathbf{a}_{t,n})$ for $\mathbf{a}_{t,n}$ by searching lookup tables $\mathcal{M}_{R,n}$ and $\mathcal{M}_{P,n}$, respectively \cite{Powell07}. Although the lookup tables can be approximated as kernel functions or neural networks\cite{SB98} to accelerate convergence, this is not the focus of this paper.
Each entry in a lookup table (which can be regarded as a matrix) denotes an action value of executing an action $\mathbf{a}_{t,n}$ in a state $\mathbf{s}_{t,n}^\prime$.
In order to reduce the space complexity of the proposed algorithm, at the beginning (where $t=0$ and the AP has not received any signal, such that all the $M$ elements in $\mathbf{s}_{0,n}$ are equal to $\sigma_n^2$), the lookup tables are initialized as row vectors, i.e. $\mathcal{M}_{R,n} = q_{0,R} \cdot \mathbf{1}_{1\times \text{card}(\mathcal{A}_n)}$ and $\mathcal{M}_{P,n} = q_{0,P} \cdot \mathbf{1}_{1\times \text{card}(\mathcal{A}_n)}$, where each element (which is nothing else than an action value) corresponds to the initial state vector $\mathbf{s}_{0,n}^\prime$ and a potential action.
Moreover, the state space $\mathcal{S}_n$ for mini-slot $n$ in the GF transmission is initialized as $\mathcal{S}_n = \mathbf{s}_{0,n}^\prime$.
In any arbitrary following timeslot $t^\prime$, if a new state is observed i.e. $\|\mathbf{s}_{t^\prime,n}^\prime - \mathbf{s}_{n}^\prime\|/\|\mathbf{s}_{n}^\prime\| > \eta \quad \forall \mathbf{s}_{n}^\prime \in \mathcal{S}_n$, the new state $\mathbf{s}_{t^\prime,n}^\prime$ is added into the state space i.e.
\begin{equation}
\label{EqStateSpaceComb}
\mathcal{S}_n = \mathcal{S}_n \cup \mathbf{s}_{t^\prime,n}^\prime \,.
\end{equation}
In the meanwhile,
\begin{IEEEeqnarray}{c}\label{EqLookupTableCombIniVec}
\mathcal{M}_{R,n} = \text{cat}\big(\mathcal{M}_{R,n}, q_{0,R} \cdot \mathbf{1}_{1\times \text{card}(\mathcal{A}_n)}\big), \nonumber  \\
\mathcal{M}_{P,n} = \text{cat}\big(\mathcal{M}_{P,n}, q_{0,P} \cdot \mathbf{1}_{1\times \text{card}(\mathcal{A}_n)}\big)\,.
\end{IEEEeqnarray}

In order to optimize the resource configuration action $\mathbf{a}_n^\star$ for multiple objectives, we define a vector-valued function $\mathbf{q}_n(\mathbf{s}_{t,n}^\prime, \mathbf{a}_{t,n}) = [\mathcal{Q}_{R,n}(\mathbf{s}_{t,n}^\prime, \mathbf{a}_{t,n}), \mathcal{Q}_{P,n}(\mathbf{s}_{t,n}^\prime, \mathbf{a}_{t,n})]^T$, so as to form a synthetic objective function. Given a quantized spectrum utilization state $\mathbf{s}_{t,n}^\prime$, in order to maximize the weighted-sum objective, we can obtain the optimized action $\mathbf{a}_n^\star$ by maximizing a synthetic objective function \cite{NT05}
\begin{equation}
\label{EqOptimAction}
\mathbf{a}_n^\star = \arg \max_{\mathbf{a}_n \in \mathcal{A}_n} \left\{ \mathbf{w}^T \mathbf{q}_n(\mathbf{s}_{t,n}^\prime, \mathbf{a}_n) \right\}\,.
\end{equation}
Motivated by the standard $\epsilon$-greedy method\cite{SB98}, to avoid always achieving the local optimum, (\ref{EqOptimAction}) is performed with a probability of $1-\epsilon$ for $\epsilon \in (0,1)$.
That is, given a random number $\epsilon_x \sim \mathcal{U}(0,1)$, if $\epsilon_x \geq \epsilon$, perform (\ref{EqOptimAction}).
Nevertheless, the standard $\epsilon$-greedy method can make the dynamic resource configuration strictly suboptimal. This due to that even if $\pi_n$ can converge over iterations, randomly selecting an action $\mathbf{a}_n \in \mathcal{A}_n$ can make UEs transmit at frequencies occupied by the interferers with a certain probability.
In order to handle this issue, we propose a novel exploration strategy: for $\epsilon_x < \epsilon$, $\mathbf{a}_n^\star$ is achieved by randomly selecting an action $\mathbf{a}_n$ from the set $\{\mathbf{a}_n | \mathbf{w}^T \mathbf{q}_n (\mathbf{s}_{n}^\prime, \mathbf{a}_n) = w_R \cdot q_{0,R} + w_P \cdot q_{0,P} , \forall \mathbf{a}_n \in \mathcal{A}_n\}$, for a given $\mathbf{s}_{n}^\prime \in \mathcal{S}_n$.
At timeslot $t$, this $\mathbf{s}_{n}^\prime$ means $\mathbf{s}_{t,n}^\prime$ which is a quantized version of $\mathbf{s}_{t,n}$. The aforementioned set essentially collects the actions that have not been explored in state $\mathbf{s}_{t,n}$.
The action executed in mini-slot $n$ of timeslot $t$ is $\mathbf{a}_{t,n} = \mathbf{a}_n^\star$. In the meanwhile, the immediate reward can be achieved as $\mathbf{r}_{t,n}$. Thereby, in mini-slot $n$ of timeslot $t$, a spectrum utilization state vector $\mathbf{s}_{t+1, n}$ can be observed, and the quantized state can be achieved as $\mathbf{s}_{t+1, n}^\prime$.

Then, by respectively defining $\kappa_q$ and $\kappa_r$ as the learning rates for updating the estimated action values $\mathbf{q}_n^{(t+1)}(\mathbf{s}_{t,n}^\prime, \mathbf{a}_{t,n})$ and the estimated average reward $\bar{\mathbf{r}}_n^{(t+1)}$, the update can be obtained as
\begin{IEEEeqnarray}{rl}\label{EqActValUpdate}
\mathbf{q}_n^{(t+1)}(\mathbf{s}_{t,n}^\prime, \mathbf{a}_{t,n})  ={}  & \mathbf{q}_n^{(t)}(\mathbf{s}_{t,n}^\prime, \mathbf{a}_{t,n}) (1 - \kappa_q) + \nonumber \\
& \kappa_q ( \mathbf{r}_{t,n} \! - \! \bar{\mathbf{r}}_n^{(t)} \! + \! \mathbf{q}_n^{(t)}\!(\mathbf{s}_{t+1,n}^\prime, \mathbf{a}_n^\star ) )
\end{IEEEeqnarray}
and
\begin{IEEEeqnarray}{rl}\label{EqAvgRewardUpdate}
\bar{\mathbf{r}}_n^{(t+1)} {} = {} & \bar{\mathbf{r}}_n^{(t)} (1 - \kappa_r) + \kappa_r ( \mathbf{r}_{t,n} \! + \! \mathbf{q}_n^{(t)}(\mathbf{s}_{t+1,n}^\prime, \mathbf{a}_n^\star ) \nonumber\\
& {} - {} \mathbf{q}_n^{(t)}(\mathbf{s}_{t,n}^\prime, \mathbf{a}_{t,n}) ),
\end{IEEEeqnarray}
where $\mathbf{a}_n^\star = \arg \max_{\mathbf{a}_n} \left\{ \mathbf{w}^T \mathbf{q}_n(\mathbf{s}_{t+1,n}^\prime, \mathbf{a}_n) \right\}$. It is noteworthy that only in the case where $\mathbf{a}_{t,n}$ is \emph{not} generated by the exploration strategy, can the estimated average reward $\bar{\mathbf{r}}_n^{(t+1)}$ be updated\cite{Mahadevan96}.
The proposed MORL-based DRC scheme is summarized in Algorithm \ref{AlgMORL}, where the processes in the for-loops can be performed in parallel at each iteration. Additionally, in the presence of $w_P$ equal to zero, Algorithm \ref{AlgMORL} amounts to a scheme based on the conventional R-learning\cite{SB98}.

\section{Performance Evaluation}
\begin{figure}[t]
\centering
\includegraphics[width = 3.2in]{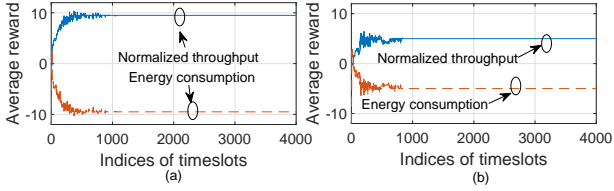}
\caption{Estimated average reward as a function of timeslots with LoS channels. (a) $w_R = 1$ and $w_P=0.5$. (b) $w_R = 1$ and $w_P=0.93$.}
\label{Fig_AvgRewards_ChanPL}
\end{figure}
\begin{figure}[t]
\centering
\includegraphics[width = 3.2in]{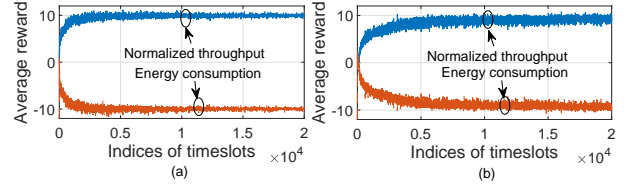}
\caption{Estimated average reward as a function of timeslots with Rayleigh fading channels. (a) $w_R = 1$ and $w_P=0.5$. (b) $w_R = 1$ and $w_P=0.93$.}
\label{Fig_AvgRewards_ChanRayleigh}
\end{figure}
\begin{figure}[t]
\centering
\includegraphics[width = 3.2in]{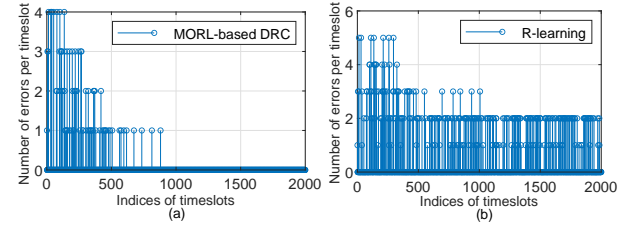}
\caption{Number of decision errors per timeslot (as a function of timeslots) yielded by various schemes with LoS channels. (a) The MORL-based DRC. (b) The R-learning\cite{SB98}.}
\label{Fig_NumErrorPerTimeslot}
\end{figure}
In the simulations, a conventional R-learning-based DRC scheme, designated as R-learning, is exploited as a baseline,
where the R-learning\cite{SB98} features a single objective and is capable of solving an MDP with an average cost.
The R-learning aims at maximizing $\bar{R}_n\, \forall n$, and the corresponding immediate reward is $R_{t,n}$ (where for fairness we assume that for an arbitrary $i$, $\alpha_{i, m, n, t}$ can be equal to 0 $\forall m$. Hence, it is possible that all UEs do not transmit at a certain time interval).
In the simulations, $K_U = 2$, $K_I = 1$, $N = 6$, $M = 6$, $p_{U,i} = 0.1\,W$ and $p_{I,i} = 0.2\,W$. The weight vector $\mathbf{w}$ is set as $w_R = 1$ and $w_P=0.5$, unless otherwise stated. Moreover, the spectrum utilization (which is observed by the AP and used as the input of the MORL-based DRC) of the interferer is periodic, and the time-frequency channels occupied by the interferer are shown in Fig. \ref{Fig_GFtimeDomainStruct}. It can be seen from Fig. \ref{Fig_GFtimeDomainStruct} that for $K_U = 2$ the maximum achievable normalized throughput per timeslot is equal to 10. The frequency-domain channels are supposed to be i.i.d., and the average signal-power attenuation w.r.t. large-scale fading is normalized as 1. We consider two types of wireless channels: Rayleigh fading channels (where $h_{i, m, l} \sim \mathcal{CN}(0,1)$ and $g_{j, m, l} \sim \mathcal{CN}(0,1)$), and Line-of-Sight (LoS) channels (where $h_{i, m, l}$ and $g_{j, m, l}$ are normalized as 1). The former is related to terrestrial communications, while the latter is related to air-ground communications.

Fig. \ref{Fig_AvgRewards_ChanPL} depicts the estimated average reward (achieved by the MORL-based DRC) as a function of timeslots in the presence of LoS channels. It is shown that for $w_R = 1$ and $w_P=0.5$, the average reward $\bar{R}$ w.r.t. normalized throughput converges to 9.5, slightly less than the maximum achievable value 10. The slight difference comes from the weight $w_P$ for the objective of energy saving. As $w_P$ increases to 0.93, when the average rewards converge, $\bar{R}$ reduces to 5, while the average reward $\bar{P}$ w.r.t. energy consumption (or energy saving) increases to -5.
Similarly, Fig. \ref{Fig_AvgRewards_ChanRayleigh} studies the convergence of the average rewards $\bar{R}$ and $\bar{P}$ in the presence of Rayleigh fading channels. The comparison of Figs. \ref{Fig_AvgRewards_ChanPL} and \ref{Fig_AvgRewards_ChanRayleigh} indicates that $\bar{R}$ and $\bar{P}$ saturate much earlier with LoS channels than Rayleigh fading channels. Furthermore, the effect of an increasing $w_P$ on $\bar{R}$ with LoS channels is more significant than that with Rayleigh channels, due to the absence of channel fluctuations.
Moreover, it can be drawn from Fig. \ref{Fig_AvgRewards_ChanRayleigh} that the convergence time of $\bar{R}$ and $\bar{P}$ scales with the weight $w_P$, although this relation is not that significant in the presence of LoS channels.

Fig. \ref{Fig_NumErrorPerTimeslot} investigates the number of decision errors per timeslot achieved by the MORL-based DRC and R-learning, where a decision error means that the time-frequency channel (related to a certain pair of $n$ and $m$) through which a certain UE transmits data is simultaneously occupied by interference. It can be seen from Fig. \ref{Fig_NumErrorPerTimeslot} that both of the number of errors achieved by the MORL-based DRC and that achieved by R-learning become steady after 1000 timeslots, as the two algorithms converges. Thus, the MORL-based DRC and the conventional R-learning achieve similar convergence time. Fig. \ref{Fig_NumErrorPerTimeslot}(b) illustrates that compared to the MORL-based DRC, the R-learning suffers from more decision errors, even if the algorithm converges.
Intrinsically, the reason lies in that no matter whether the UEs are configured to transmit in mini-slot $6$ (in Fig. \ref{Fig_GFtimeDomainStruct}) of each timeslot, the normalized throughput (i.e. the only immediate reward $R_{t,n}$ involved in the R-learning) always remains zero. This leads to the potential for configuring UEs to transmit even if frequency-domain resources in a mini-slot are totally occupied.

\begin{figure}[t]
\centering
\includegraphics[width = 3.2in]{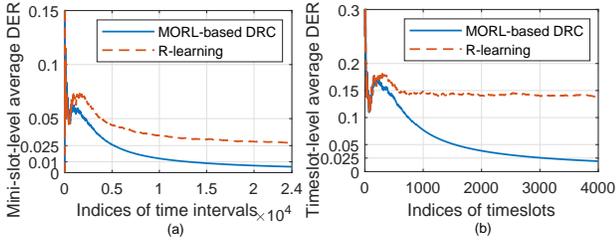}
\caption{Average Decision Error Rate (DER) achieved in the presence of LoS channels. (a) Mini-slot-level average DER. (b) Timeslot-level average DER.}
\label{Fig_DER_ChanPL}
\end{figure}
\begin{figure}[t]
\centering
\includegraphics[width = 3.2in]{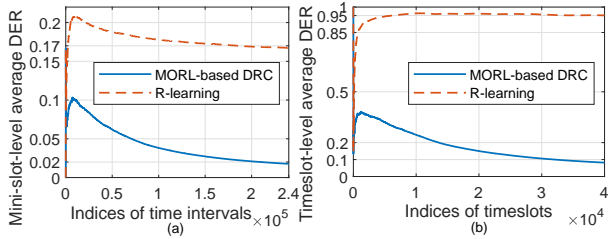}
\caption{DER achieved in the presence of Rayleigh fading channels. (a) Mini-slot-level average DER. (b) Timeslot-level average DER.}
\label{Fig_DER_ChanRayleigh}
\end{figure}
Fig. \ref{Fig_DER_ChanPL} presents average Decision Error Rate (DER) as a function of timeslots in the presence of LoS channels. The mini-slot-level/timeslot-level average DER is computed by averaging the number of mini-slots/timeslots, during which decision errors occur, over the number of elapsed mini-slots/timeslots. It is shown that the mini-slot-level and the timeslot-level average DERs achieved by the R-learning finally reach values around 0.025 and 0.15, respectively. This huge gap illustrates that although the R-learning does not always configure the UEs to transmit in mini-slot $6$ (in Fig. \ref{Fig_GFtimeDomainStruct}), the average DER performance can still be heavily degraded. On the contrary, the mini-slot-level and the timeslot-level average DERs yielded by the MORL-based DRC are less than 22\% (around 0.0055) and 13\% (around 0.0195) of those yielded by the R-learning, respectively.
%
Fig. \ref{Fig_DER_ChanRayleigh} illustrates the average DER performance with Rayleigh fading channels, where the mini-slot-level and the timeslot-level average DER achieved by the R-learning finally reaches values around 0.17 and 0.95, respectively. This observation implies that in this simulation with channel fluctuations, the R-learning configures the UEs to transmit in mini-slot $6$ in most timeslots. The comparison of Figs. \ref{Fig_DER_ChanPL} and \ref{Fig_DER_ChanRayleigh} reveals that although the average DER performance achieved by the MORL-based DRC can be degraded due to channel fluctuations, the mini-slot-level and the timeslot-level average DER  with Rayleigh fading channels can be less than $0.02$ and $0.1$ (which are less than $12\%$ of those achieved by the R-learning), respectively (while most of the decision errors occur before reaching convergent solutions).

\section{Conclusions}
\label{SecConclu}
In this paper, we have proposed a DRC scheme based on MORL for GF uplink transmissions in IoT networks. Thanks to the multi-objective framework, the proposed scheme is able to not only pre-allocate time-frequency resources for UEs without the knowledge of interference pattern, but also guarantee energy-efficient transmission.
It is shown that in the presence of Rayleigh fading channels, the average DER achieved by the MORL-based DRC can be even less than $12\%$ of that yielded by the R-learning-based method, especially when frequency-domain channels are totally interfered in a time interval.
Integrating kernel/neural network-based function approximations with the framework can be studied in the future to accelerate the convergence of the algorithm.


\bibliographystyle{IEEEtran}
\bibliography{IEEEabrv,Bib_IoT_FreqSche}

\end{document}